\documentclass{pasa}%

\usepackage{graphicx}

\title[3C\,273, the First Quasar]{The Sequence of Events that led to the 1963 Publications in {\em Nature} of 3C\,273, the First Quasar and the First Extragalactic Radio Jet}
\author[Hazard et al.]{Cyril Hazard$^{1,2}$, David Jauncey$^{3,4,}$\thanks{e-mail: David.Jauncey@csiro.au}, W. M. Goss$^{5}$ and David Herald$^{6}$
\affil{$^1$University of Pittsburgh, Pittsburgh, PA, USA}%
\affil{$^2$Institute of Astronomy, Cambridge, UK}
\affil{$^3$CSIRO Astronomy \& Space Science, NSW, Australia}
\affil{$^4$Research School of Astronomy \& Astrophysics, Australian National University, Australia}
\affil{$^5$National Radio Astronomy Observatory, Socorro, NM, USA}
\affil{$^6$International Occultation Timing Association, NSW, Australia}
}%

\jid{PASA}
\doi{10.1017/pas.\the\year.xxx}
\jyear{\the\year}

\usepackage{aas_macros}
\usepackage{hyperref} 
\hypersetup{colorlinks,citecolor=blue,linkcolor=blue,urlcolor=blue}

\begin{document}

\begin{frontmatter}
\maketitle

\begin{abstract}
We have undertaken a detailed investigation, based on the available evidence, of the sequence of events that led to the historical discovery of the first quasar, 3C\,273.
\end{abstract}

\end{frontmatter}

\section{INTRODUCTION }

The pages of {\em Nature} for 1963 March 16 carried four short publications; the first two announced the precise position, structure, and optical identification and redshift measurement of the radio source 3C\,273, the first quasar \citep{Hazard1963,Schmidt1963}. These were followed by the recognition of the very blue colour of the 3C\,273 identification, as well as the identification of the H$\alpha$ line in the near-infrared \citep{Oke1963}. Then followed the recognition of the high redshift and blue colour in the existing spectrum of the radio source 3C\,48 \citep{Greenstein1963}. These discoveries irreversibly changed our understanding of the Universe, and at the same time gave 3C\,273 an iconic place in extragalactic astronomy.

Fifty-five years later we have undertaken a detailed examination of the circumstances surrounding the observations that led to the publications in {\em Nature} in 1963 by Cyril Hazard et al., and Maarten Schmidt, of the precise radio position and structure and the optical identification of 3C\,273.

The 3C\,273 discovery was an international effort made by an Englishman working in Australia and a Dutchman working in the United States. It was made using two of the world's largest telescopes at that time, the Parkes 210~ft.\ (64~m) and the Palomar 200~inch (5~m). We are indeed fortunate in our investigations that fifty years ago national and international communication was essentially by letter. Many of these letters have survived and remain accessible through archives and collections available today, and a number are reproduced here. This is a more complete version of our earlier published account \citep{Hazard2015} and now includes additional information recently uncovered from such archival sources. These sources include the National Archives of Australia, the University of Sydney Archives, the CSIRO Archives, as well as Cyril Hazard's personal papers. It is important to delve deeply into these historical archives to understand the often complex histories, as professional historians often have a difficult time sorting it all out, because so much depends on the technical details. The historians see only the finished product, and what actually took place behind the scenes, in terms of the development and exchange of scientific ideas and concepts, can be hidden from them.

\section{BACKGROUND}
The first extensive radio sky surveys had revealed the presence of two distinct source populations. The strong and often extended Class~I sources were generally concentrated close to the Galactic plane, while the remaining Class~II sources were distributed uniformly over the sky \citep{Mills1952}. One of these bright and extended Class~I sources was identified in 1949 as `one of the minor discrete sources of galactic radio-frequency noise---Taurus-A' the Crab Nebula \citep{Bolton1949}. The radio position had been determined with the cliff interferometer in New Zealand with quoted errors of $\pm 30$ seconds of time in Right Ascension (RA) and $\pm 8$' in Declination. This position was found to enclose the optical nebula NGC\,1952. It was also noted that the effective temperature at 100~Mc/s (100 MHz) of two million degrees, suggested that non-thermal processes were responsible for the radio emission \citep{Bolton1949}.

The positions for two other discrete radio sources were also measured, and these were identified with NGC\,5128, Centaurus~A, and NGC\,4486, Virgo~A \citep{Bolton1949Nat}. They suggested that both were Galactic nebulae, although both were generally considered as extragalactic. Soon afterwards, \cite{Brown1950} detected the radio emission from the Great Nebula in Andromeda, M31, which supported arguments for the extragalactic nature of both Centaurus~A and Virgo~A. But the majority of the Class~II discrete sources were not associated with bright optical objects. The identification of the intense radio source, Cygnus~A, with an unusual and distant extragalactic object, possibly a pair of colliding galaxies \citep{Baade1954} made it clear that the Class~II sources could also be extragalactic objects at great distances \citep[R.~Minkowski, in][p.~188]{1953Obs}.

It was thus a major task to establish the nature of these Class~II sources. For this, accurate and reliable radio positions and structures, preferably both with arcsecond precision, were needed in order to accurately locate their optical counterparts amongst the plethora of faint stars and galaxies. In the late 1950s and early 1960s radio position measurements with this precision were not available, e.g.\ the 3C catalogue \citep{Edge1959,Bennett1962}. In California, John Bolton had developed the Owens Valley Radio Observatory Interferometer yielding radio positions with $\sim 15$ arcsec accuracy \citep{Read1963,Maltby1963}.

Hazard had worked with Hanbury Brown, observing with the Jodrell Bank pencil-beam 218~ft reflector at 158.5~Mc/s., and had detected emission from a number of late-type spirals. He suggested that the majority of the fainter Class~II sources appear to be distributed isotropically and hence they were likely to be extragalactic \citep[C.~Hazard, in][p.~193]{1953Obs}.

In his PhD thesis, `An Investigation of the Extra-Galactic Radio Frequency Emissions' submitted in 1955, after a careful analysis of their survey data, he concluded `that the majority of the Class II sources so far observed are probably extra-galactic objects.'

After completing National Service, Hazard returned to Jodrell Bank in 1957, where he started exploring the capability of lunar occultations with the Jodrell Bank 250~ft (76~m) telescope which had only been brought into operation in late 1957 \citep[c.f.][]{Lovell1959}. While occultations had been used previously, Hazard was specifically interested in their ability to determine radio positions and structure with arcsecond precision. Hazard undertook his first successful radio occultation observations at Jodrell Bank in an investigation of the radio position and structure of 3C\,212. The radio position was determined with a precision of 3 arcsec, the most accurate radio position measurement made up to that time. 3C\,212 was slightly extended on the scale of $\sim 10$ arcsec, but the optical field showed no prominent visual object \citep{Hazard1961,Hazard1962}.

Most importantly, these observations demonstrated the power of lunar occultations in enabling the measurement of radio positions with arcsecond precision. Such measurements were made in the optical reference frame thereby allowing reliable optical identifications. Without the capability for measuring accurate optical positions, Hazard was not then able to make the identification. At that time such capabilities were scarce even at Caltech (California Institute of Technology). Later, with access to the Palomar Sky Survey, Hazard identified 3C\,212 with a 19.5 magnitude `stellar-like' object \citep{Hazard1964}.

In 1961, together with John Davis, he joined the Chatterton Astronomy Department at the University of Sydney to supervise the construction of the Brown-Twiss Intensity Interferometer at Narrabri, NSW \citep{Brown1967}, which was being led by Professor Hanbury Brown at the University of Manchester and Jodrell Bank. Hazard also continued his occultation work with the full support of Professor Harry Messel, Head of the School of Physics, although Hanbury Brown had expressed some reservations about Hazard doing so (Messel personal communication).

Maarten Schmidt studied with Jan Oort in Leiden where he earned his PhD in 1956. He immigrated to the United States in 1959 to begin work at Caltech. In 1961, on Rudolph Minkowski's retirement, he took over the radio source optical identification program. Using the Mount Palomar 200-inch telescope for his spectroscopy, Schmidt found that the first three objects initially `identified', on the basis of the Owens Valley radio positions, were in fact misidentifications \citep{Schmidt1983}. However, his observations in May 1962 of the optical object associated with 3C\,286 exhibited one broad emission line at 5170\,\AA, and he noted that this line was not observed in any other astronomical object, including the spectrum of the `star' identified with 3C\,48 \citep{Schmidt1962}. \citet{Shklovsky1963} suggested that, given the lack of other strong emission lines in Schmidt's spectrum, the likely identification was with the MgII 2798\,\AA ~line. The redshift of 3C\,286 was later determined by \citet{Oke1965} as 0.849, thereby confirming Shklovsky's suggestion. In May 1962 Schmidt also obtained a spectrum of a galaxy mistakenly identified at the time with 3C\,273 \citep{Schmidt1983}.

\section{THE PARKES OCCULTATIONS}

Following a talk at Sydney University on his 3C\,212 and Jodrell Bank occultation experiences, Hazard was invited by Joseph Pawsey, Assistant Chief of the CSIRO Division of Radiophysics, to use the Parkes telescope as a Guest Observer, and to continue his occultation observations \citep{Hoyle1981}. Hazard had been in Australia at Sydney University since early 1961, but Hanbury Brown, still associated with Manchester University, did not arrive until a year later. While visiting Parkes for the observing, Hazard was to be well supported by Brian Mackey and John Shimmins as well as by the Parkes Director, John Bolton. In a letter to Hanbury Brown in January 1962, Hazard mentions that `a lunar occultation programme is being arranged there', and then notes that `the feuding between the University and C.S.I.R.O. at the higher levels and with which you are probably familiar (and if not will be) doesn't seem to be affecting relations between the groups seriously'. The state of these `higher level' relations became important after the final identification and publications, as we show later in the story.

Hazard's motivation for the occultation work was to provide the precise, arcsecond radio positions for high Galactic latitude Class II radio sources that were essential for establishing their nature. Hazard had noted that the strong Class~II radio source 3C\,273 would be occulted by the Moon several times during 1962 and 1963, a once in twenty year occurrence. Moreover, these occultations would be accessible at CSIRO's newly completed Parkes 210 foot radio telescope. The importance of these occultations was that they provided the only technique then available that was capable of yielding a precise radio position with arcsecond accuracy, and that the measured radio position would be in the optical reference frame.

In 1962, the year of these observations, Maarten Schmidt was working on the program of optical identification and spectroscopy of the optical objects identified with radio sources \citep{Schmidt1962} based on radio positions measured with the Owens Valley interferometer \citep[e.g.][]{Read1963}. In this he was ably supported by Tom Matthews who made the essential optical position measurements \citep[e.g.,][]{Matthews1962}.

\subsection{The Parkes Occultation of 1962 May 15}

Three occultations of 3C\,273 were predicted for 1962. The first was for May 15, when only the reappearance would be visible, the second was for August 5, when both the disappearance and reappearance were to be observed. The third was for October 26, when only the disappearance would be visible. In 1962 Parkes could not track the Moon, only the source.

\begin{figure}
\begin{center}
\includegraphics[width=\columnwidth]{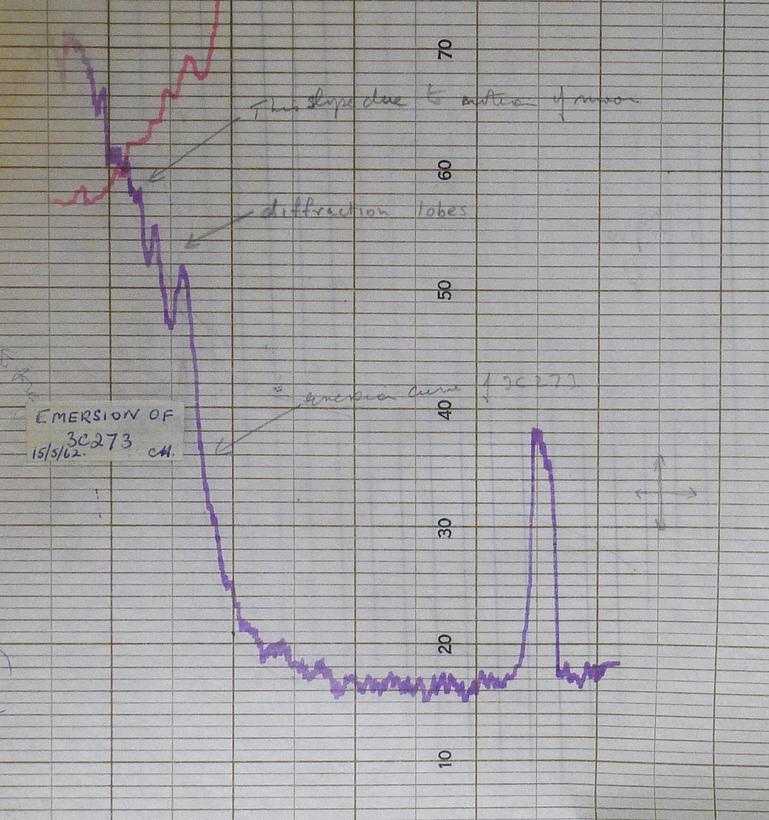}
\caption{The 1962 May 15 Parkes occultation record at 410~MHz. The rapid increase in the received noise level can be clearly seen, as can the diffraction fringes. Time increases from right to left. The plot is taken from Hazard's personal records.}\label{figure1}
\end{center}
\end{figure}

The first of the occultation observations at Parkes took place on 1962 May 15 at 410~MHz, when only the reappearance was predicted to be accessible. For clarity we use the terms `disappearance' and `reappearance' instead of the terms `immersion' and `emersion'. The reappearance occurred as the Moon was rising and entering the field of view of the telescope, with the telescope stationary at the elevation limit. The diffraction fringes can be seen clearly in Figure~\ref{figure1}, but the record was complicated by the presence of the rapid increase in level as the Moon entered the beam of the telescope. Despite these problems, what interested Hazard was the clear presence of well-defined diffraction fringes. This was the first time that such fringes had been seen and they clearly identified the presence of a component of diameter no more than $\sim 2$ arcseconds. Such a small component implied that the necessary arcsecond position would be achievable. Unfortunately, the rapid increase in level as the Moon entered the beam made the record unsuitable for a more detailed analysis \citep{Hazard1963}. But it did make clear the importance of the upcoming August 5 occultation for which both disappearance and reappearance would be accessible at Parkes.

\subsection{The Parkes Occultation of 1962 August 5}
Before the scheduled 1962 August 5 observations, Bolton was concerned about the predicted low elevation for the reappearance and undertook significant `surgery' on the telescope. As John Shimmins recalled later `We simply removed the safety stops from operation for that observation. Just to be certain John Bolton took a grinder and cut away some of the housings of the zenith angle bearings. During the occultation the rim of the dish was practically touching the ground.' (quote from John Shimmins in Beyond Southern Skies \citep[][pp.~230]{Robertson1992}). The utmost care was taken by Bolton for these observations. Observers were stationed around the site with Malcolm Sinclair stationed in the master equatorial room on the Parkes telescope with a small optical telescope to independently observe the Moon, all roads leading to the observatory were closed, unnecessary electrical equipment on site was shut down and duplicate records of the occultation were to be made.

Because of Bolton's adjustments to the Parkes telescope made to allow observations below the 30 degree elevation limit, we looked to determine the telescope elevation at the time of the observed reappearance, 09:05:45.7 UT, given by \citet{Hazard1963}. We did this in two ways, firstly to use the observed reappearance time to calculate the elevation, and secondly we used the known source position to predict the time and elevation for Parkes. The first method gave an elevation of 30.13 degrees, and then adding an additional 1.5 arcminutes for ionospheric refraction at 410\,MHz, yielded a value of 30.16 degrees and a set time at 30 degrees of 09:06:36 UT. These figures were independently confirmed by our recalculating the predictions for the known position of 3C\,273 and the position of Parkes. Interestingly, if Parkes had been able to track the Moon rather than just the source, this would have added a further 2 minutes of time before 3C\,273 set.

This was something of a surprise, especially given the reports of the concern about the predicted low elevation for the 1962 August 5 reappearance, and the modifications done to the telescope in order to allow tracking at elevations below the nominal limit (e.g.\ Frank Kerr, In Serendipitous Discoveries in Radio Astronomy, 1983, p174)\nocite{Kerr1983}. Later, while examining the CSIRO Archives we found copies of the predictions sent to Pawsey by Mrs Sadler at H.M.\ Nautical Almanac, which were headed `Sydney Fleurs'. Further searching revealed the letter requesting future occultation predictions, which Pawsey had sent to the Almanac Office in 1958. At that time he had requested that the predictions be made for the CSIRO Fleurs Field Station, as construction of the Parkes Telescope had not then commenced. Pawsey's request had come as part of the research planning for the Parkes telescope. 

The differences between occultation predictions for Fleurs and for Parkes are significant, as Fleurs is 275\,km east-south-east of Parkes, two and a half degrees east in longitude and almost a degree south of Parkes. Because of the Moon's motion from west to east, the August reappearance occurred first at Parkes almost three minutes before the reappearance at Fleurs. Moreover, the elevation at Fleurs is significantly lower than the elevation at Parkes for the earlier time of the observed reappearance there. It would seem that much of the telescope modifications made before the 1962 August 5 occultation were in fact unnecessary, as the observations were complete before the telescope reached its 30 degree elevation limit, and well before Frank Kerr's estimated twenty eight and a half degrees.

However, aside from the Almanac Office predictions there is a simple and direct means of establishing accurately when 3C\,273 reaches the setting elevation limit at the Parkes telescope. This is for the telescope to track 3C\,273 into the limit beforehand, and then to use the `4 minutes a day' solar-sidereal shift to determine when it will set on 1962 August 5. This technique does not rely on knowing the exact position of either 3C\,273 or the Parkes telescope. Such an exercise could have revealed that 3C\,273 reappearance would remain visible at Parkes minutes before the telescope reached its 30 degree elevation limit, and so would have alerted the Parkes astronomers. Moreover, the 48 and 144 arcminute half-power beam widths of the telescope at 410 and 136 MHz respectively, as were used in August, would easily allow the observations an additional offset of 5 to 10 arcmin or more before causing serious problems. Also, allowing the telescope to continue tracking in azimuth while in the elevation limit would extend this further. As noted above, the recalculation of the predictions using Parkes instead of Fleurs, demonstrate that the reappearance took place significantly above, although close to the telescope elevation limit.

After the 1962 August and October occultation observations, the lesson regarding the use of the Fleurs rather than Parkes predictions had been learned at Parkes. On November 1, 1962, Bolton wrote to Mrs Sadler at H.M.\ Nautical Almanac Office and asked that the predictions in future be made for the Parkes Radio Telescope, not for the Fleurs Field Station. Mrs Sadler replied on November 22 `We shall in future calculate the predictions for occultations of radio sources for Parkes instead of Fleurs.'

Hazard was using predictions that he had requested himself from the Almanac Office, and that these predictions were for Parkes, not for Fleurs. As the telescope operations for the occultations were all undertaken by the CSIRO staff, he was at that time unaware that they were using Fleurs predictions. He later needed the Parkes predicted occultation details in order to undertake his reductions of the structure and positions of 3C\,273.

\begin{figure*}
\begin{center}
\includegraphics[width=0.99\textwidth]{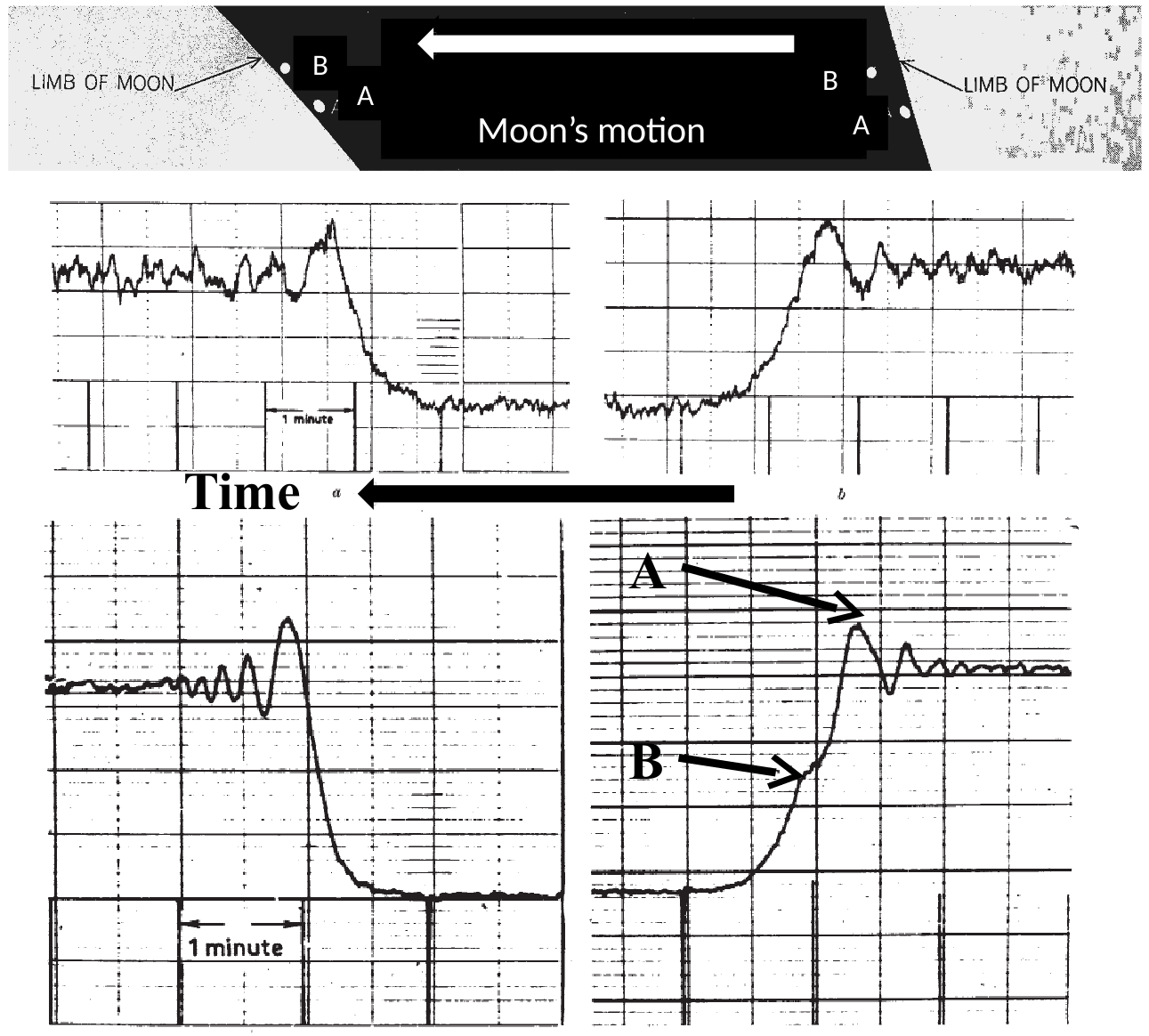}
\caption{The August 5, 1962 disappearance and reappearance records at 136~MHz, centre, and 410~MHz. bottom, taken from \citet{Hazard1963}. Note that time increases from right to left, and that the Moon is also moving from right to left. The top panel shows the positions of source components A and B relative to the limb of the Moon at disappearance and reappearance.}\label{figure2}
\end{center}
\end{figure*}

The 1962 August 5 occultation was undertaken with concentric feeds at 410 and 136~MHz where the telescope half-power beamwidths are 48 and 144 arcminutes, respectively. Both disappearance and reappearance were observed and the disappearance record revealed the presence of two components, A and B, in the source. The disappearance on the right and the reappearance records on the left, at 136 and 410 MHz are plotted in Figure~\ref{figure2}; time increases from right to left, and the motion of the Moon is also from right to left.

The basic source structure of 3C\,273 is apparent from figure 2. At 410~MHz the disappearance record shows that 3C\,273 is a double source with both components no more than a few arcseconds in size. There is a 30 second separation between the disappearance of component A, the brighter one at 410 MHz, and the fainter component B. The 410~MHz reappearance record shows that the components are aligned so that both reappear together along the 45 degrees position angle of the Moon's limb at the reappearance. However, the disappearance record at 136 MHz shows no evidence for component B seen at 410 MHz. At 136 MHz component B is weak or non-existent compared to component A.

On 1962 August 9, four days after the occultation, a telegram, found in the Narrabri interferometer archives, was sent to Rudolph Minkowski on Hazard's behalf, stating succinctly `Occultation success source double components three seconds probably elliptical separated ten seconds position being calculated'. The Moon moves across the sidereal sky at one third of an arcsecond per second, hence the 30 second time difference between the disappearance times of the two components at 410~MHz in figure 2, yields a 10 arcsecond projected component separation. 

Hazard started work on the reduction of the 1962 August 5 occultation soon after returning from Parkes, knowing that the October 26 occultation was still to come. Starting with the 136~MHz records was straightforward as the observed disappearance and reappearance indicated essentially a single source. The procedures he had developed for the occultation of 3C\,212 were fresh in his mind as the 3C\,212 paper had only recently been received by MNRAS on May 10 \citep{Hazard1962}. The presence of strong fringes in both the disappearance and reappearance at 136 MHz showed that the source was compact on a scale of a few arcseconds, the scale of the first Fresnel zone at the Moon's limb. The 136 MHz position was then determined following the procedures used for 3C\,212.  The uncertainties in this position were expected to be $\sim 5$'', and arose from the lack of knowledge of the details of the Moon's limb, uncertainties in the precise difference between UT and Ephemeris time, uncertainties in the position of the Parkes telescope, and also uncertainties in the precession calculation. However, this position was sufficient for the preparations for the October occultation.

At 410 MHz the situation differs from that at 136 MHz. At 410 MHz the record of the disappearance showed fringes from two close compact components separated by 30 seconds of time, so the diffraction patterns for the first disappearing component, now known as component A, is confused, and that for the second, now known as B, hardly visible. The reappearance shows both components reappearing at the same time, thereby merging the two diffraction patterns. A further difficulty is that no clear angular sizes can be determined from either the disappearance record or the reappearance record because of the confusion between the fringes. At this early stage of the analysis an `average' was used to determine the position of the `centroid', that is, the mean of the disappearance times for the two components.

Hazard's initial analysis of the two 1962 August 5 occultation records produced two estimates of the centroid position of 3C\,273 separated by 10 arcseconds, (a) at 136 MHz and (b) at 410 MHz. With this analysis in hand, Hazard crossed the Sydney University campus to visit John Bolton at the CSIRO Division of Radiophysics, which was then also on campus, to pass on these initial results. While there, Minkowski, who was already working with Bolton, joined the group. As Minkowski had previously been advised of the 3C\,273 occultations by Hazard's telegram, he had with him a Palomar Polaroid of the 3C\,273 field, and because of a previous Caltech misidentification, he also had a measured optical position for a reference object on the Polaroid. It was a simple matter for Minkowski using a ruler to determine the approximate optical positions for the occultation positions of (a) and (b). Hazard's recollection of Minkowski's response was `It looks like an edge-on spiral, and we know that spiral galaxies are not strong radio emitters'. Figure~\ref{figure4} shows that it was what we now know is the `jet' that Minkowski had `identified'.

On 1962 August 20, soon after Hazard's visit, Bolton wrote to Maarten Schmidt (section reproduced below in Figure~\ref{figure3}) 
about a separate potential Parkes identification. He also asked, as a by the way, `Would you also pass on to Tom (Matthews) the following position for 3C\,273', but gave no reference as to its origin. He also added a P.S.\ saying `The position of 3C\,273 is subject to arithmetic errors as yet undetected!'

\begin{figure}[h]
\begin{center}
\noindent \fbox{\includegraphics[width=0.98\columnwidth]{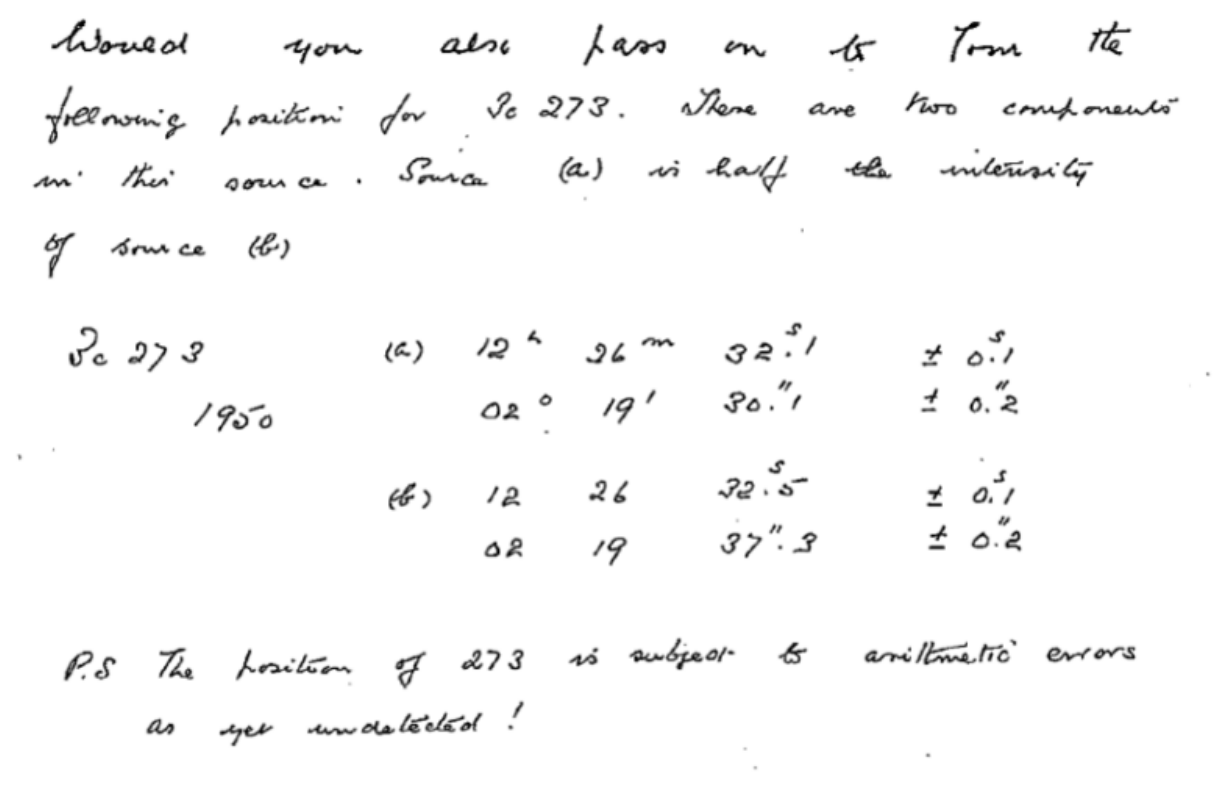}}
\caption{Section of the letter Bolton wrote to Maarten Schmidt on 1962 August 20.}\label{figure3}
\end{center}
\end{figure}

The contents of the above letter raise several important questions, both by statements in the letter as well as by what is not stated in the letter. First of all, Bolton wrote `..following position for 3C\,273', but he then interprets the two separate occultation positions given to him by Hazard, as the positions of the two components, rather than two separate estimates of the centroid position, where the quoted position for source (a) is the position of the 136~MHz occultation, and source (b) the estimated centroid position at 410~MHz.

Secondly, Bolton's quoted uncertainties appear quite arbitrary. In both (a) and (b) the claimed uncertainty in right ascension is $\pm 0.^s1$, or 1.5 arcseconds, whereas the quoted Declination error is 0.2 arcseconds, almost eight times smaller. These numbers were not the result of Hazard's analysis. The final quoted errors given by Hazard et al., in their 1963 {\em Nature} paper for the actual separate components A and B, based on both the August and October occultations are, for A, $\pm 0.^s03$ and $\pm 1.''5$, and for B, $\pm 0.^s02$ and $\pm 0.''5$ in RA and Declination respectively.

What does not appear in Bolton's letter is any mention of the person responsible for deriving these numbers. Moreover, this letter was not shown to Hazard, and neither did it let Schmidt know that Hazard was the person originally responsible for the analysis. Thus there was no opportunity for Hazard to correct Bolton's misinterpretation of the two separate positions. Nor was there an opportunity for Hazard to update Schmidt as the analysis developed, especially after the October occultation. Similarly, there was there no opportunity for Schmidt to communicate his findings with Hazard. 

\begin{figure*}
\begin{center}
\includegraphics[width=0.9\textwidth]{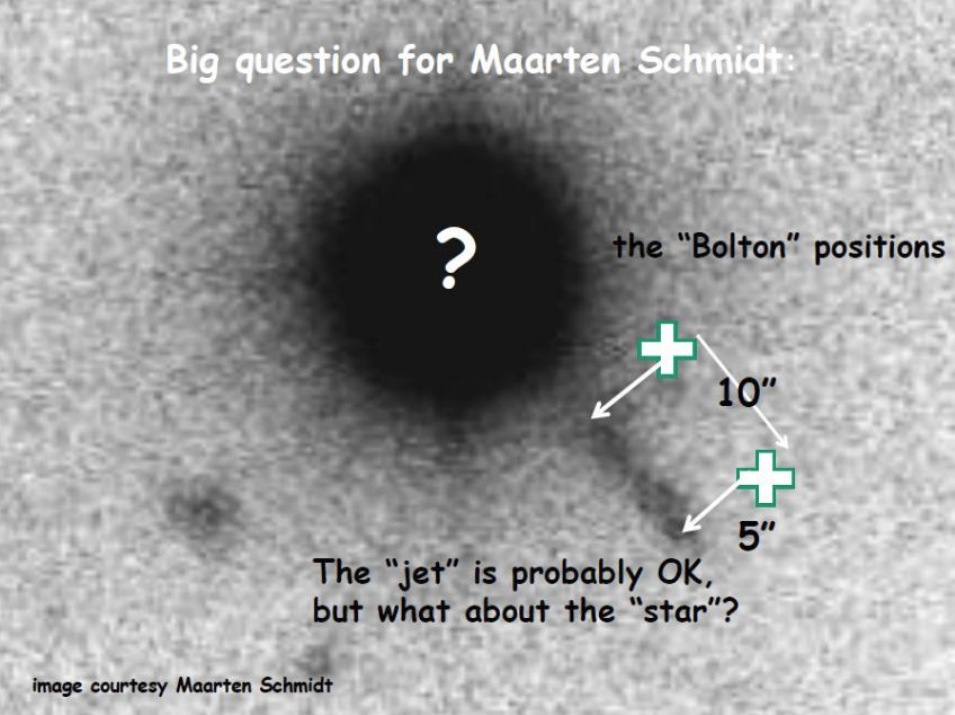}
\caption{Palomar 200-inch image of the field in the vicinity of 3C\,273. Superimposed are the `Bolton' positions sent to Schmidt on 1962 August 20. North is up and East to the left. (Background image courtesy of Maarten Schmidt)}\label{figure4}
\end{center}
\end{figure*}

Figure~\ref{figure4} shows the Sky Survey image with the `Bolton' positions superimposed. It clearly suggests that the optical identification for 3C\,273 was most likely what we now know as the jet; the radio source is at exactly the same $45^{\circ}$ position angle as the jet and is the same length, only displaced laterally by about 5 arcseconds. Such a coincidence would otherwise be very unlikely. This jet is just what Minkowski had showed Hazard and Bolton and which he had described as an `edge-on spiral'. 

A question for Schmidt, however, was what role does the bright, 13th magnitude `star' play in the identification? Was it merely a random foreground star, or alternatively, was it somehow associated with the `jet'? His assessment, given the Bolton positions, was that it was a random foreground star. At the time this was not pressing because for the next four months, 3C\,273 was up only in daylight hours, and so was inaccessible at Palomar.

\subsection{The Parkes Occultation of 1962 October 26}
\begin{figure*}
\begin{center}
\includegraphics[width=0.99\textwidth]{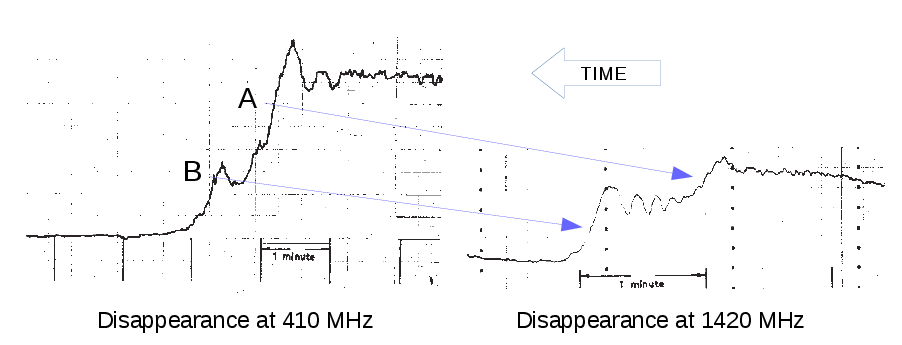}
\caption{The 3C\,273 disappearance records of the 1962 October 26 Parkes occultation observations at position angle 83 degrees, from \cite{Hazard1963}. Again time increases from right to left as does the right ascension.}\label{figure5}
\end{center}
\end{figure*}

The 1962 October 26 disappearance-only observations were undertaken successfully, this time at 410 and 1420 MHz, the standard concentric feed configuration in use at that time for the Parkes Sky Survey. The October occultation results are shown in Figure~\ref{figure5}. At 410 MHz both components show clear diffraction fringes, indicating angular sizes of 1--2 arcseconds for both. At 1420 MHz, where the Fresnel scale is much smaller, component A shows at best a small bump, while B exhibits strong fringes. Component B is significantly smaller than A at the sub-arcsecond scale. A detailed independent analysis of these observations was later undertaken \citep{Scheuer1965} showing that both components at 410 MHz have diameters less than 2 arcseconds, while at 1420 MHz component B is less than 0.5 arcseconds in diameter.

Most interesting is that component B is stronger than component A at 1420 MHz. The spectral index, $\alpha$, defined for frequency $\nu$ by flux density $S \propto \nu^{\alpha}$, of the more compact component B between 410 and 1420 MHz is flat at $\alpha = 0.0$, whereas that of A is steep at $\alpha = -0.9$. The 1962 August record shows component A totally dominates at 136 MHz with $\sim 95$\% of the total flux density. Here was the first concrete evidence for an inverted-spectrum, sub-arcsecond component, B, in a radio source. The low-frequency surveys of the 1950s had preferentially found the steep-spectrum sources. It would be the later high-frequency surveys that would uncover for the first time many more of these compact, flat-spectrum sources.

\begin{figure*}
\begin{center}
\includegraphics[width=0.99\textwidth]{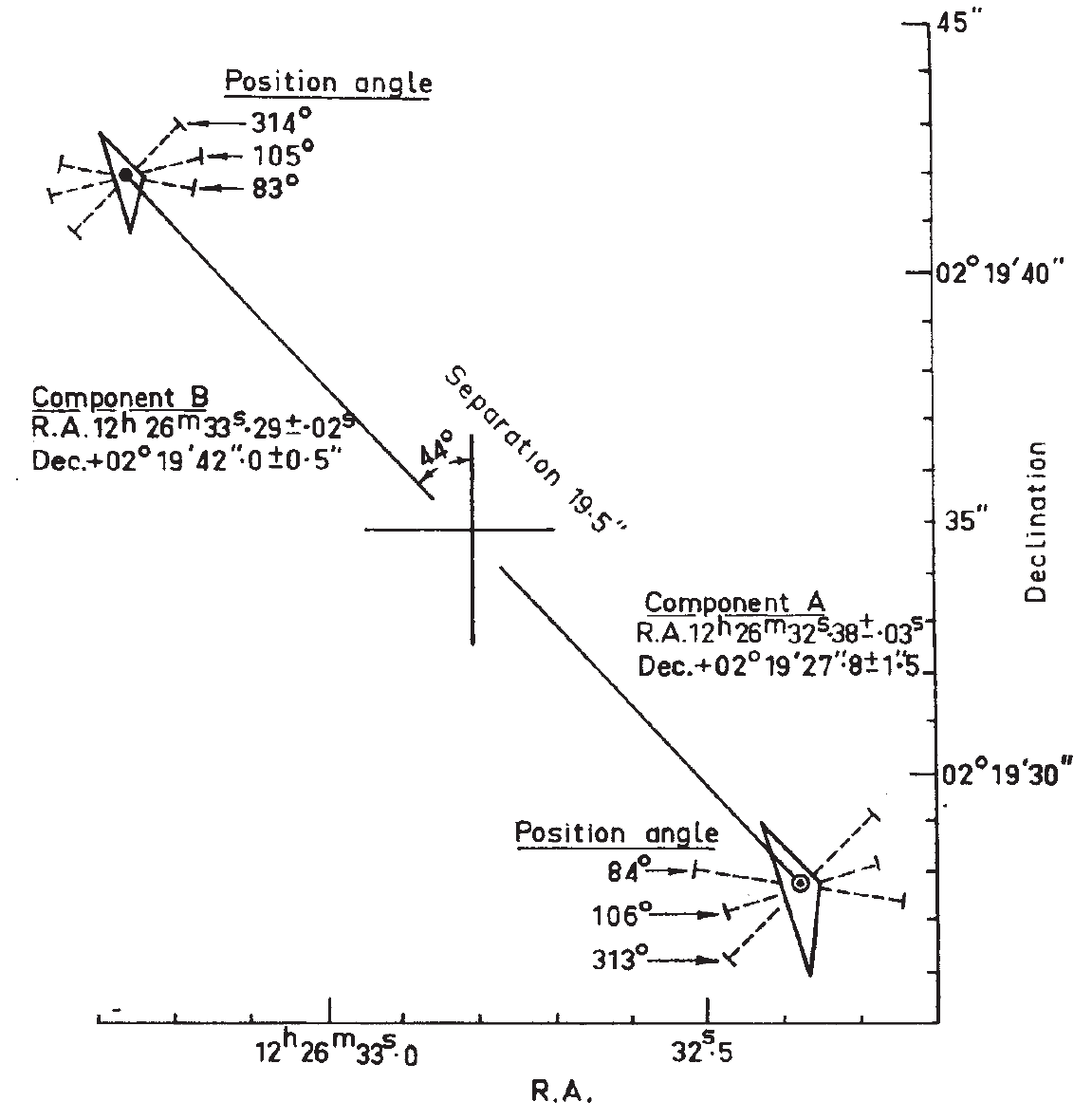}
\caption{The occultation structure and positions from both the 1962 August 5 and October 26 Parkes occultations of 3C\,273, from \citet{Hazard1963}.}\label{figure6}
\end{center}
\end{figure*}

For the 1962 August and October occultations, Hazard's 3C 273 analysis had revealed a `core-jet' structure with `normal' steep-spectrum component A and an inverted-spectrum, very compact component B. The final image and positional information was determined by Hazard in collaboration with Nicholson, and the detailed arcsecond resolution image is given here in Figure~\ref{figure6}, taken from \citet{Hazard1963}.
Their determination of the final 3C\,273 position and structure took place following the October occultation, and its evolution can be followed in an exchange of letters between Hazard and Nicholson. On December 17 Hazard wrote to Nicholson letting him know that he was `sending a letter to {\em Nature} with the preliminary results'. The last exchange took place in response to Hazard's letter of December 17, after which Nicholson made the final arcsecond adjustments. 

After the October occultation and in parallel with these developments, there had been an exchange of strongly worded letters, found in the Sydney University Archive, between Professor Messel and Professor Hanbury Brown, regarding the division of Hazard's efforts between the Narrabri Intensity Interferometer and his continued involvement in the Parkes occultation program. Construction of the interferometer at Narrabri was running into a number of difficulties with the equipment as delivered from the UK. Hanbury Brown wished to keep Hazard and John Davis, the Sydney University staff members involved with the intensity interferometer, to remain on-site and fully occupied with the interferometer. 

On the other hand, Messel could see the deadlock that was unfolding at Narrabri and felt that Hazard and Davis could be more usefully employed, whenever possible, doing their own research work rather than involving themselves fully with problems that were the responsibility of the UK contractors. On 1962 November 15, Messel wrote to Hanbury `I stress again that this is neither your, nor Davis's, nor Hazard's responsibility. In view of this I have requested Davis and Hazard not to concern themselves for the time being with the mechanical difficulties of the instrument and to devote the majority of their time to their own scientific projects.' 

Hanbury Brown was not at all happy, `I expect to be consulted before the only two academic staff are instructed not to help me.' But after remonstrating with Messel he conceded `I intend to continue work on the reflectors until we have diagnosed all the troubles and can put all their faults clearly to the engineers.' The difficulty for Hanbury Brown was that Davis and Hazard were employees of Sydney University School of Physics, and thus were Messel's responsibility. Hanbury Brown on the other hand, while having moved to Narrabri to supervise the construction of the intensity interferometer, was at this time on leave of absence from Manchester University, so had no direct control of Davis and Hazard. Davis and Hazard remained at Narrabri and continued with their own scientific projects, but also continued working on the interferometer. The first observations with the Intensity Interferometer, undertaken in 1963 July-August, were of the angular diameter of the star $\alpha$-Lyrae \citep{Brown1964}, and in 1964 Hanbury resigned his position with the University of Manchester and took on a professorship at Sydney University.

As pointed out earlier, as Hazard had not seen the Bolton letter of 1962 August 20, to Schmidt, there was no direct line of communication between Hazard and Schmidt. Moreover, as Schmidt had to wait until the year's end to attempt spectroscopy with the 200-inch telescope, what communication there had been was indirect, via intermediaries. John Whiteoak, from Radiophysics, had moved to Mt Wilson in early November 1962 as a Carnegie Fellow, to discuss the extension of the Palomar Observatory Sky Survey (POSS; \citealt{Minkowski1963}) to more southerly declinations. While there he discussed the occultation results with Tom Matthews, as John Whiteoak had seen the occultation records in Australia. Whiteoak noted that Matthews had mentioned a `star and a funny jet' as the only objects in the field on the  Palomar 48-inch (1.2m) Schmidt plate. In a long letter to Bolton discussing the Palomar Extension, Whiteoak mentions as an afterthought that the `current Caltech thinking' is that the potential 3C\,273 identification is with a star and a strange jet.

Hazard was prompted to finally write to Maarten Schmidt when he heard from Jim Roberts that Schmidt had succeeded in identifying 3C\,273. Roberts had been in the US at the `Physics of Nonthermal Radio Sources' meeting in New York, where he actually showed pictures of the Parkes August and October occultation plots in the discussion following Alan Moffet's presentation \citep[][pages 35 and 36]{Moffet1964}. Roberts had worked at Caltech several years previously, and had been passing through Pasadena visiting John Whiteoak and Venkatraman Radhakrishnan over Christmas 1962. On 1963 January 30, Hazard wrote to Schmidt with the correct occultation structure and positions for A and B, and suggested a joint publication. On 1963 January 26, Bolton wrote to Schmidt also giving the occultation positions.

\section{SCHMIDT'S 200-INCH SPECTRUM}
In the meantime, on the nights of 1962 December 27 and 29, Schmidt had undertaken spectroscopy at the 200-inch, his first opportunity to do so since receiving Bolton's August 20 letter. He was faced with a practical dilemma. Given the Bolton positions, and given that no other bright star had been proposed as a radio source identification, he assumed that the bright 13th magnitude `star' was merely a confusing foreground very bright star. To obtain a spectrum of the faint jet, which he saw as by far the most likely identification, he had to place the slit of the spectrograph along the extended jet for a long exposure, even with the 200-inch. In doing so it was inevitable that the bright confusing star 10 arcseconds away, would spill over into any spectrum of the jet he would obtain.

To offset this, Maarten Schmidt had decided to first obtain a spectrum of this bright star. His first attempt on the night of 1962 December 27 was overexposed. Although he had the world's largest optical telescope at his disposal, at that time his detector was a small photographic plate with efficiency no more than a few percent, so exposure times were not always easy to determine. On the night of December 29 he managed to obtain a spectrum of the bright `star' that was not overexposed, and which showed some faint emission lines, but with no obvious explanation in terms any expected stellar lines.

There it stayed until 1963 February 6th, as outlined in Schmidt's paper presented to the American Philosophical Society \citep{Schmidt2011}. He wrote:
{\bf
\begin{quotation}
`It happened on 6 February 1963. In response to Hazard's letter I decided to have another look at the spectra... For reasons that I don't remember I tried to construct an energy-level diagram. When the energy levels did not come out regularly spaced, I was annoyed... To check on the regularity of the observed lines, I decided to compare them with the Balmer lines of hydrogen... Specifically, I took for each line in 3C\,273 the ratio of its wavelength over the wavelength of the nearest Balmer line. The first ratio was 1.16, the second was... also 1.16.

It suddenly struck me that I might be seeing a redshift. When the third and fourth ratios were also close to 1.16, it was abundantly clear that I was seeing in 3C\,273 a redshifted Balmer spectrum.'
\end{quotation}
}

\begin{figure*}
\begin{center}
\includegraphics[width=0.9\textwidth]{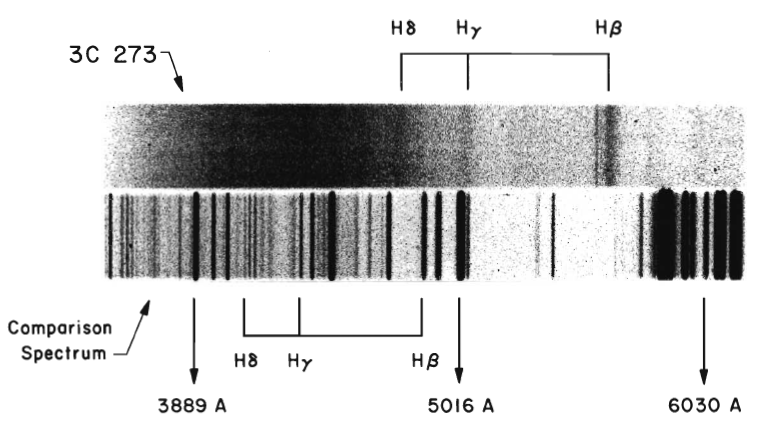}
\caption{Maarten Schmidt's 200-inch spectrum taken on the night of 1962 December 29, with the identified Balmer lines at a redshift of 0.16 (figure courtesy of Maarten Schmidt).}\label{figure7}
\end{center}
\end{figure*}

It was indeed fortunate that the 3C\,273 redshift was small enough that the Balmer series was there and would be readily recognised, as can be clearly seen in Figure~\ref{figure7}.

\begin{figure*}
\begin{center}
\includegraphics[width=0.99\textwidth]{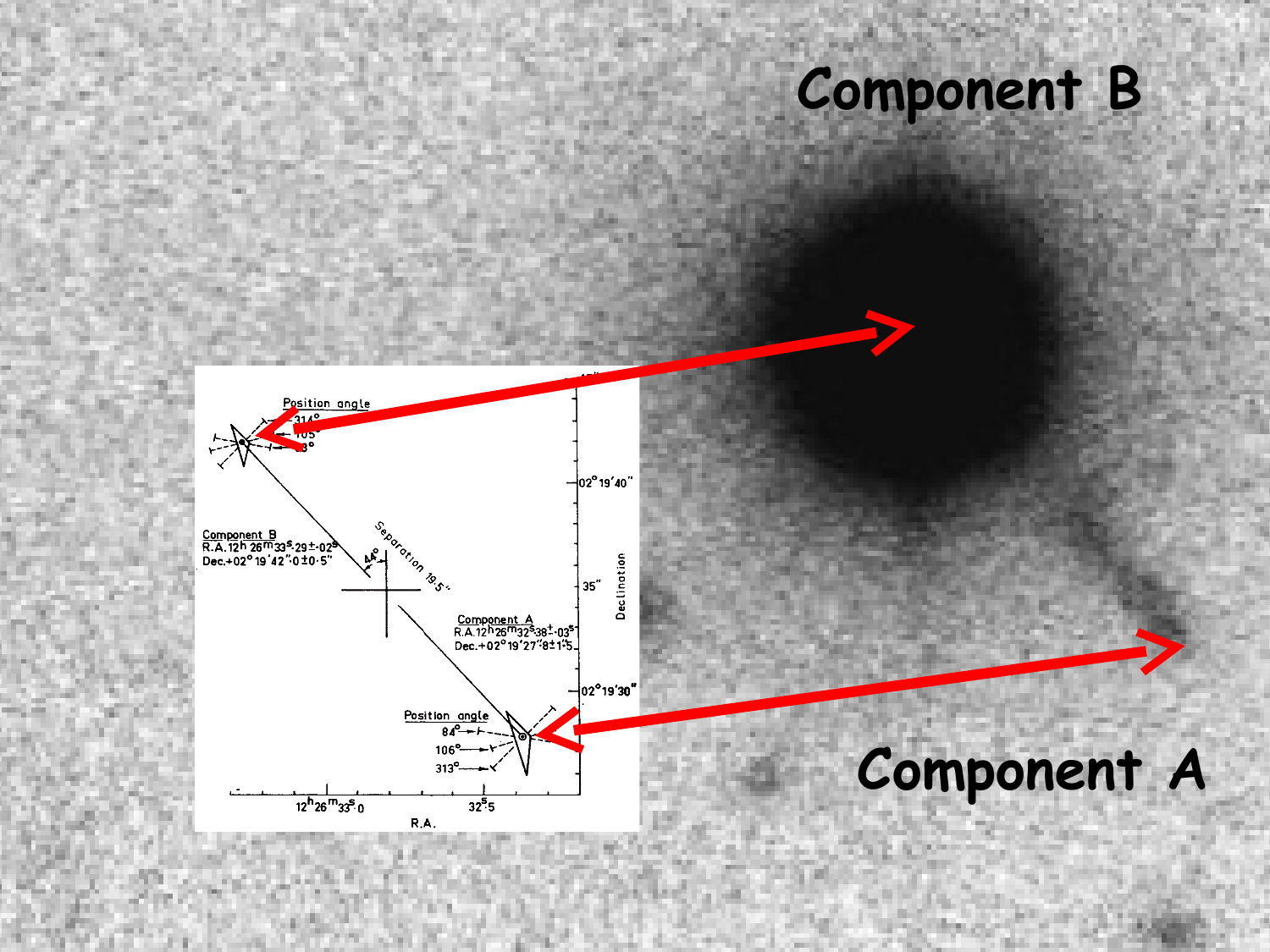}
\caption{The culmination of the match between the Parkes radio position and the Palomar image of 3C\,273.}\label{figure8}
\end{center}
\end{figure*}

In discussions since, Maarten Schmidt has raised the question as to who had actually identified 3C\,273. It is clear that it was only when faced with Hazard's final radio structure and position as shown in Figure~\ref{figure8}, that it was finally possible to make the correct identification. The bright star and the tip of the faint jet were co-incident with the radio components B and A respectively. The consecutive papers in {\em Nature} of 1963 March 16 followed. With the discovery of the first quasar our knowledge and understanding of the Universe changed forever.

3C\,273 became:
\begin{itemize}
  \item the first Quasar
  \item the first radio and optical Jet
  \item the first inverted spectrum radio source
  \item the first sub-arcsecond radio position
  \item the first sub-arcsecond radio structure
  \item the first radio-optical reference frame tie
  \item the first optical and radio variable extragalactic source
  \item the first black hole.
\end{itemize}

Following the success of the 3C\,273 observations Hazard continued with the Parkes occultation program, and a year later had added another 15 occultation observations of 8 sources at 136, 410 and 1420 MHz \citep{Hazard1964}. He concluded that, of the extragalactic sources, the stellar type comprise 25--30\% of the total extragalactic identifications. Hazard also noted of the identifications, that `many of them appear to be characterised by having a large fraction of their emission concentrated in a region of small angular size ($<1$ arcsecond), and the occultation observations yield the required diameter information at the same time as the accurate positions required for their identification.' Hazard moved to Arecibo in 1964 to continue his occultation observations now using the 1,000 ft reflector \citep{Hazard1967}.

Schmidt continued his very effective and fruitful spectroscopic program on the Palomar 200-inch telescope. His 1966 study of nine 3C quasars \citep{Schmidt1966IAU} showed all were stellar in appearance and also showed an ultra-violet excess, while three were optically variable. Of interest to radio astronomers searching for quasars, Schmidt found that the associated radio sources did not uniformly stand out by any one radio property. For example, their measured radio angular sizes varied between less than 1 arcsecond for 3C48, through to $\sim 1$ arcminute for 3C\,47.

The optical identification process changed dramatically with the optical identification now of both 3C 273 and 3C 48 as very blue stellar objects. Faint blue stellar objects are orders of magnitude less frequent on the Palomar Sky Survey than faint red objects, which meant that to identify radio quasars, the radio positional accuracy required for reliable identifications was significantly reduced, and positions with larger errors, as high as 20 or even 30 arcseconds, became sufficient to determine reliable blue stellar object identifications. Many more quasar identifications soon followed. At Parkes, Bolton realised the consequences and the Parkes identification work expanded rapidly especially with the Whiteoak southern extension to the Palomar Sky Survey \citep{Bolton1965}. At Caltech, Wyndham proposed 10 possible quasar identifications as well as confirming nine others \citep{Wyndham1965}. With access to the Palomar 200-inch, Schmidt was well placed to provide the confirming redshifts from more occultations, the Owens Valley interferometer and Parkes \citep{Schmidt1966}. 

An important component of these identifications was that they also provided an increasing number of compact radio position calibrators across the sky, calibrators that also came with accurate arcsecond quality optical positions from the Palomar plates and prints. By 1966, 69\% of all 3CR sources with Galactic latitude $> 15$ degrees had been optically identified, of which 30\% were with quasi-stellar sources \citep{Wyndham1966}. Another important outcome from the occultations and identifications was that for the first time the radio and optical astronomers began to work together much more closely, both nationally and internationally.

\section{A PRECISE RADIO OCCULTATION POSITION FOR 3C 273}
As a check on the coincidence of the radio and optical positions for 3C\,273B, Hazard undertook a reanalysis of the occultation positions, together with new measurements of the optical position. \citet{Hazard1971} selected only those occultation curves for which the signal-to-noise ratio and the separation of the A and B components permitted a timing accuracy of better than 0.1 seconds, corresponding to an uncertainty in the location of the Moon's mean limb of $\leq 0.3$ arcseconds. These were the 1962 August 5 and October 26 Parkes observations at 410 MHz, as well as the 1963 March 11 Parkes observations at 2650 MHz, plus the Arecibo 430 MHz observations of 1965 June 7 \citep{Hazard1966}. Their careful reanalysis took account of the accurate lunar ephemeris j=2, and used recent analysis of stellar occultations by \citet{Morrison1969}. 

The optical positions for 3C\,273B were determined at both the Royal Greenwich Observatory and at Cambridge. The close agreement between both the radio and optical positions confirmed the accuracy of each group's measurements, and further showed that both exhibited positions accurate to 0.2 arcseconds \citep{Hazard1971}. Their final analysis gave (in B1950 coordinates):\\

\begin{tabular}{ll}
Radio & RA(B1950) = $12^{\rm h}26^{\rm m}33^{\rm s}.253 \pm 0^{\rm s}.014$ \\ 
      & Dec.(B1950) = $02^{\circ}19'43''.47 \pm 0''.3$\\
Optical & RA(B1950) = $12^{\rm h}26^{\rm m}33^{\rm s}.246 \pm 0^{\rm s}.01$ \\
& Dec.(B1950) = $02^{\circ}19'43''.38 \pm 0''.1$.\\
\end{tabular}\\

These occultation positions were then used by many to align the radio and optical reference frames \citep[c.f.][]{Johnston1983}.

We have reanalysed the occultation data from the same experiments as used by \cite{Hazard1971}, in order to ascertain what really can be achieved for 3C\,273 with the 1960s lunar occultation data from Parkes and Arecibo. At the time of these observations the height of the lunar limb was poorly known, as the available charts were based on measurements from photographs. In our analysis the lunar limb is derived from altimeter measurements from the Lunar Orbiter Laser Altimeter, LOLA, on NASA's Lunar Reconnaissance Orbiter spacecraft \citep{Smith2009}. This provides a dataset of surface heights around the lunar limb at intervals of 59 metres, with a height resolution of 0.5 metres. These are a factor of $\sim 10$ more precise than the earlier Watts Tables \citep{Watts1963} used by \citet{Hazard1971}. The LOLA data reveals remarkably smooth limb profiles for both the 1962 August 5 and October 26 occultations. The values from our reanalysis yielded smaller errors, and gave a B1950.0 position of:\\

\begin{tabular}{l}
RA(B1950) = $12^{\rm h}26^{\rm m}33^{\rm s}.242 \pm 0^{\rm s}.005$\\
Dec.(B1950) = $02^{\circ}19'43''.46 \pm 0''.2$.\\
\end{tabular}\\

There is excellent agreement between all three measurements, confirming the \citet{Hazard1971} error estimates. 
%
%

\section{POSTSCRIPT: THE AFTERMATH}

While the four key publications in {\em Nature} \citep{Hazard1963,Schmidt1963,Oke1963,Greenstein1963}, changed astrophysics so dramatically, another drama was about to unfold. In writing the 3C\,273 paper, Hazard had arranged for the CSIRO Division of Radiophysics to prepare and submit it as a Letter to {\em Nature}. The file for this publication, RPP 779, the 779th paper to proceed through the Radiophysics publication office, originally listed the authorship as:\\

\noindent C.~Hazard\\
School of Physics, University of Sydney\\
M.~B.~Mackey\\
A.~J.~Shimmins\\
Radiophysics Laboratory, CSIRO, Sydney,\\
Australia\\

However, before final submission, this was changed with the following revision.
\\

\noindent\includegraphics[width=0.99\columnwidth]{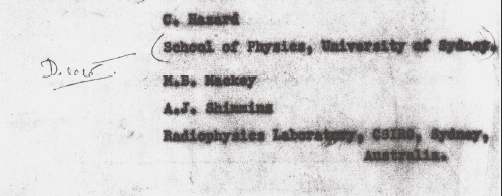}
\\

It is not known who made this revision, but the net result was that the typed version submitted to the journal was:
\\

\noindent\includegraphics[width=0.99\columnwidth]{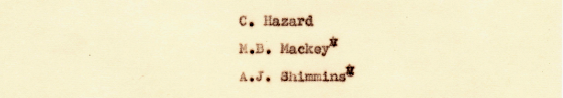}\\
\noindent\includegraphics[width=0.99\columnwidth]{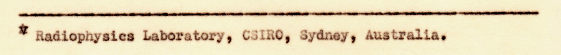}
\\

As we have seen, the addresses appeared in {\em Nature} as:
\\

\noindent\includegraphics[width=0.99\columnwidth]{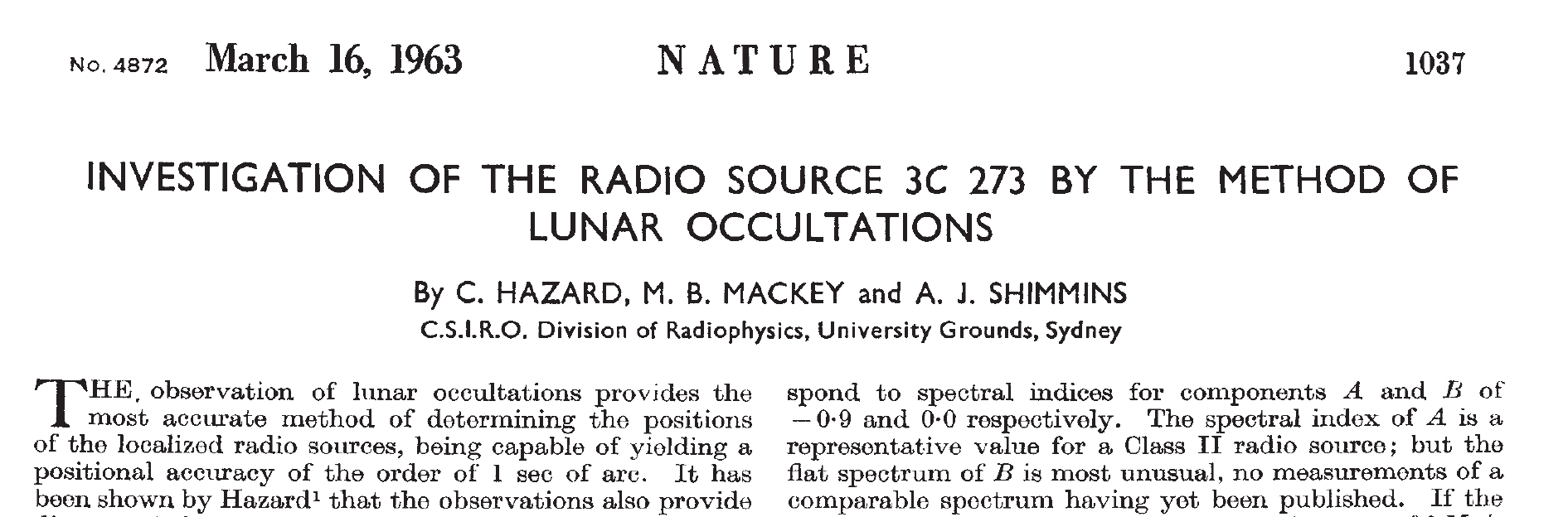}
\\

When the mistake was pointed out to the journal by John Bolton (letter White to Messel, 1963 April 16), {\em Nature} apologised and it noted that part of the change could have arisen because the original {\it Letter} became a full paper, and the following appeared in the 1963 April 6 issue.
\\

\noindent\includegraphics[width=0.99\columnwidth]{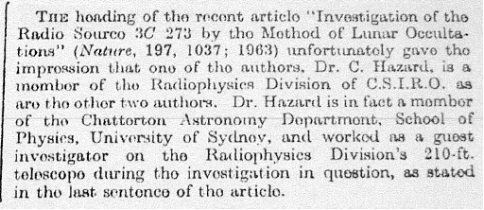}
\\

Hazard initiated the Parkes 3C\,273 occultation program with the specific intention of using occultations as a tool for determining the precise radio position and structure of 3C\,273. He was clearly its principal investigator, as well as being responsible for analysing the occultation curves, and it was he who wrote the paper to {\em Nature}. The manuscript was typed and readied for publication by CSIRO. In the process, the archive RPP779 shows that CSIRO added a paragraph without consulting Hazard. And it was CSIRO who sent on the manuscript to {\em Nature}, without sending a copy of the submitted version to Hazard, although a copy was sent to Maarten Schmidt as well as to CSIRO Head Office. But no copies were recorded as being sent to Hazard or to anyone at Sydney University.

It was CSIRO who changed the details of Hazard's affiliation in the manuscript that was sent to {\em Nature}. Regardless of the reasons for these manuscript changes, it was clear that the changes came from CSIRO. The original list of affiliations was perfectly clear. However, this last change to the submitted manuscript could be described, at best, as ambiguous, specifically in regard to Hazard's affiliation. The process of changing Hazard's affiliation has also been discussed by \citet{Haynes1993} who concluded `Through an innocent editorial mistake, the affiliation of the first author was wrongly attributed.' {\em Nature} made a simple mistake, and they promptly followed this with an apology.

By itself this should not necessarily caused a serious problem between Sydney University and CSIRO. The simplest and least confrontational response from CSIRO to Sydney University would have been an apology, exactly as did {\em Nature}.

However, the CSIRO publicity machine went into high gear with the Australian press three days after the March 16 publication in {\em Nature}. Claims appeared of a `major CSIRO discovery made by CSIRO scientists', and headlines like `CSIRO Pinpoints new ``star''' as appeared in the Daily Telegraph of March 19 1963. Sydney University was not mentioned, and Hazard, when mentioned at all, was referred to as `Dr.\ P.\ Hazzard, a guest investigator from the Jodrell Bank Observatory, England'. This was most unfortunate and provided considerable fuel to the already fragile high-level relations between Sydney University and CSIRO, as Hazard had noted earlier in a letter to Hanbury Brown.

Messel was very upset at this turn of events and on 1963 March 20 wrote a strongly worded letter to Sir Frederick White, the CSIRO Chairman, which is now available through the National Archives of Australia. In it Messel says how pleased he was with Hazard's results, but how upset he was to see CSIRO apparently claiming the discovery for themselves, and noting `The next I hear in regard to this matter is the press announcement in Canberra made by officers of the Division of Radiophysics and I enclose the relevant newspaper clippings.' He further states `It would seem that if anybody was to have made any press announcements it should have been the School of Physics of the University of Sydney and not the C.S.I.R.O.'

Again, the whole misadventure could easily have been diffused had Sir Frederick, in his March 27 reply to Messel, simply apologised, as did {\em Nature}. But this did not come to pass. After an exchange of letters between Sir Frederick White and the Chief of the Radiophysics Division Dr E. G. (Taffy) Bowen, dated March 27, April 4 and April 10, the CSIRO attitude hardened significantly. 

The high level relations between the two organisations deteriorated further. In response to Messel, Sir Frederick had at no point made any comment about the last step in presenting the lack of affiliation in the manuscript that was, as we now know, sent to the journal. The turning of a clear set of addresses into the ambiguous version shown above was apparently never admitted at the time by CSIRO, and no copy of the actual submission to {\em Nature} had ever been sent to Hazard or to anyone else at Sydney University.

But Sir Frederick did make it clear to Messel that his `colleagues at Parkes have enjoyed having Hazard work with them and I hope that, if he wishes to continue, he will feel free to do so.' Thus although the deterioration in the high level relations continued, at the working level, Hazard continued his occultation observations at Parkes until he left Sydney University in mid 1964, with further outstanding successes.

We will allow Professor Harry Messel the last word: `It was a dirty and miserable affair which soured relations for years. Thankfully the years have healed most of the wound and relations with CSIRO are 100\% and so many of our former students are proudly with them. Cheers. Have fun and keep well. HM' (Harry Messel private communication to D. Jauncey 2013 February 18). Messel passed away July 8, 2015.

\section{IMPACT OF THE DISCOVERY}

Our knowledge and understanding of the Universe changed forever with the consecutive papers in {\em Nature} of 1963 March 16, with the discovery of the first quasar 3C\,273. This was due to Hazard's precise radio position from the Parkes occultations and Maarten Schmidt's subsequent optical identification and redshift determination with the Palomar 200-inch. The occultations provided the first sub-arcsecond radio position and structure, and the optical identification with the distant quasar provided a precise radio-optical frame tie.

The discovery forged, for the first time, a closer connection between radio and optical astronomers, both nationally and internationally. Radio and optical quasars were soon discovered in the early Universe, and they rapidly became a major tool of cosmology.
These connections spread rapidly with the increasing wavelength coverage that soon became available. 

As a result, quasars are now a familiar term, and knowledge of them is taken for granted. Their discovery, however, was time consuming and complex. The sequence of events that led to the publications about the first quasar, 3C\,273, is documented here through the detailed study of letters, archives, scientific papers, early telegrams, and personal papers. The archives also have given us an insight into the institutional politics at play at that time, of which the astronomers were not fully aware. But it is of primary importance that science has an accurate record of the observations of those astronomers, Cyril Hazard and Maarten Schmidt, over 50 years ago. Their discoveries dramatically changed the way scientists view the Universe, and shape our understanding of it.

\begin{acknowledgements}
We wish to thank Maarten Schmidt for many valuable discussions. We would also like to thank Martin Harwit for his encouragement, John Reynolds for the Parkes details, and Nyree Morrison of the Archives and Records Management Services at the Fisher Library of the University of Sydney, and to also thank the National Archives of Australia, Chester Hill Sydney. The National Radio Astronomy Observatory is a facility of the National Science Foundation operated under cooperative agreement by Associated Universities, Inc. We would like to thank Hayley Bignall for her time and patience in formatting the manuscript. We thank the referee for their careful reading of the text and thoughtful suggestions.
\end{acknowledgements}

\bibliographystyle{pasa-mnras}
\bibliography{pasa_3C273}

\end{document}